\documentclass{article}
\usepackage{mathrsfs}
\usepackage{arxiv}
\usepackage[utf8]{inputenc} 
\usepackage[T1]{fontenc}    
\usepackage{hyperref}       
\usepackage{url}            
\usepackage{booktabs}       
\usepackage{amsfonts}       
\usepackage{microtype}      
\usepackage{graphicx}
\usepackage{amsmath}        
\usepackage{algorithm}
\usepackage{algpseudocode}
\graphicspath{ {./} }

\title{A Bayesian Model Updating Framework for Systems Under Hybrid Uncertainties via Probability Integral Transform and Maximum Mean Discrepancy}

\author{
Shijie Zhong \\
School of Power and Energy\\
Northwestern Polytechnical University\\
Xi'an, Shanxi 710129 \\
\texttt{zhongsj@mail.nwpu.edu.cn} \\
\And
Jiangfeng Fu \\
School of Power and Energy\\
Northwestern Polytechnical University\\
Xi'an, Shanxi 710129 \\
\texttt{fjf@nwpu.edu.cn}
}

\begin{document}
	\maketitle

\begin{abstract}
Model updating under hybrid uncertainty is challenging because aleatory input variability makes the simulator output a probability distribution rather than a scalar, rendering the likelihood analytically intractable. Existing Approximate Bayesian Computation (ABC) methods typically employ nested Monte Carlo sampling, where aleatory samples are redrawn for each epistemic parameter evaluation, introducing sampling noise into the discrepancy and consequently affecting posterior inference and model evidence. This paper eliminates this resampling noise by construction. The probability integral transform (PIT) converts the stochastic simulator into a deterministic map of distribution-free latent variables and epistemic parameters. By freezing a set of stratified quantile particles, the resulting discrepancy becomes a deterministic, sampling-noise-free function of the unknown parameters. Transitional Markov Chain Monte Carlo (TMCMC) is then employed for posterior inference and model evidence estimation. The framework is validated on a two-dimensional benchmark, a high-dimensional transient oscillator, and Subproblem A of the NASA Langley Multidisciplinary Uncertainty Quantification Challenge. The complete Bayesian analysis is achieved in approximately half a minute on a standard desktop workstation.

\end{abstract}

\section{Introduction}
\label{sec:intro}
Computational simulations are widely used to design, analyze, and maintain complex engineering systems \cite{bi2023stochastic}. However, differences between simulation models and physical experiments are unavoidable. Model updating, also referred to as model calibration, is commonly used to reduce these differences by adjusting uncertain model parameters using experimental data \cite{oberkampf2004challenge, zarei2023indoor}.

Uncertainty is an important issue in model updating and is commonly classified into aleatory and epistemic uncertainty \cite{roy2011comprehensive}. Aleatory uncertainty represents the inherent variability of a physical process and is usually described by a prescribed probability distribution. Epistemic uncertainty results from incomplete knowledge of model parameters and can be reduced by incorporating additional experimental information \cite{kitahara2022nonparametric}. In many engineering problems, these two types of uncertainty occur simultaneously, leading to hybrid uncertainty. In this case, the simulator output is a probability distribution induced by the aleatory inputs, rather than a single deterministic value. Conventional deterministic calibration methods are not designed for this setting because they typically match individual experimental observations by estimating fixed parameter values \cite{moore2009introduction, faes2020recent}. Stochastic calibration methods instead aim to infer the epistemic parameters while accounting for the variability observed across repeated physical experiments \cite{calvi2005uncertainty}.

Several methods have been developed for this problem, but they have limitations in handling hybrid uncertainty. Approximate Bayesian Computation (ABC) \cite{beaumont2009adaptive, nakagome2013kernel} avoids direct evaluation of an intractable likelihood by defining a discrepancy between simulated and experimental observations \cite{bi2017uncertainty, kisamori2020model}. Maximum Mean Discrepancy (MMD), which measures the difference between probability distributions in a reproducing kernel Hilbert space, has therefore been used to construct such discrepancies \cite{gretton2012kernel, kitahara2022nonparametric}. Kernel Bayes' Rule (KBR) \cite{fukumizu2013kernel} and its importance-weighted variant (IW-KBR) \cite{xu2022importance} provide another likelihood-free approach based on empirical covariance operators.

For MMD-based updating, a key issue arises from the treatment of aleatory uncertainty. At each trial value of the epistemic parameters, the aleatory inputs are sampled again to generate an empirical simulator output distribution. Therefore, repeated evaluations at the same parameter value generally produce different empirical distributions and different MMD values. The resulting discrepancy is consequently a random estimate rather than a deterministic function of the parameters. In the numerical study in Section~\ref{subsec:case1}, this sampling variation is about $29\%$ of the discrepancy level. Such variation can affect the shape of the discrepancy landscape and, in turn, the posterior inference and model evidence obtained from it. KBR-based methods have a different limitation: they are formulated for a single observation and a single conditional distribution, and therefore do not directly use the repeated-trial structure of hybrid-uncertainty experiments.

This paper addresses the resampling problem in MMD-based updating by fixing the aleatory samples used during parameter evaluation. The separation of aleatory and epistemic uncertainty through an auxiliary variable is not new. It was introduced by Sankararaman and Mahadevan \cite{sankararaman2013separating} and has mainly been used for sensitivity analysis. The contribution of this paper is to use this decomposition for likelihood-free Bayesian model updating. Specifically, the probability integral transform (PIT) is used to express the aleatory input as
$
X = F_X^{-1}(\xi \mid \theta),
$
where $\xi$ is a distribution-free latent variable and $\theta$ denotes the epistemic parameters. The simulator output can then be written as
$
Y = h(\xi,\theta),
$
which is deterministic for fixed $(\xi,\theta)$. A fixed set of stratified quantile particles is used to represent $\xi$. Once these particles are fixed, the corresponding simulated output particle cloud becomes a deterministic function of $\theta$. The MMD calculated between this particle cloud and the experimental samples is therefore also deterministic with respect to $\theta$, removing the resampling noise from discrepancy evaluation.

The proposed framework uses a multi-scale MMD to measure the difference between simulated and experimental distributions. An energy-form surrogate likelihood is then constructed from the resulting discrepancy and sampled using Transitional Markov Chain Monte Carlo (TMCMC) \cite{ching2007transitional}. The tempering procedure of TMCMC also provides an estimate of the model evidence. Since the same particle representation and MMD formulation can be used for scalar and multidimensional outputs, the proposed construction does not require a separate formulation for different output dimensions.

The main contributions of this work are as follows:

1. A frozen-particle construction based on the existing PIT decomposition is introduced for likelihood-free model updating. By fixing the latent-variable particles, the empirical simulator output and its discrepancy with experimental data become deterministic functions of the epistemic parameters, eliminating the resampling noise in MMD-based likelihood evaluation.

2. A multi-scale MMD discrepancy and a discrepancy-scale-calibrated energy-form surrogate likelihood are developed to perform Bayesian updating directly in distribution space without requiring probability density estimation.

3. A TMCMC-based inference procedure with model evidence estimation is implemented and evaluated on synthetic benchmarks and Subproblem~A of the NASA Langley Multidisciplinary Uncertainty Quantification Challenge using the official black-box simulator and experimental data.

The remainder of this paper is organized as follows. Section~\ref{sec:existing} reviews existing methods and discusses their limitations. Section~\ref{sec:method} presents the PIT-MMD framework and the TMCMC inference procedure. Section~\ref{sec:case} evaluates the proposed method using synthetic and engineering benchmarks and discusses the computational cost. Section~\ref{sec:conclusion} summarizes the main findings.

\section{Existing methods and their limitations}
\label{sec:existing}

\subsection{Model updating under hybrid uncertainty}
\label{sec:setting}

In practical engineering systems, uncertainty arises from two fundamentally different sources. The first is \emph{aleatory uncertainty}, which represents the inherent variability of physical processes and is typically characterized by precise probability distributions. The second is \emph{epistemic uncertainty}, which originates from incomplete knowledge of the model parameters. The simultaneous presence of these two uncertainty types gives rise to a \emph{hybrid uncertainty} framework.

Let
\begin{equation}
	\theta \in \Theta
\end{equation}
denote an uncertain epistemic parameter belonging to an admissible parameter space $\Theta \subset \mathbb{R}^d$. Let $X$ be a system input governed by the conditional probability distribution
\begin{equation}
	X \sim p(X \mid \theta).
\end{equation}
The system response is given by
\begin{equation}
	Y = g(X).
\end{equation}

\begin{figure}[htbp]
	\centering
	\includegraphics{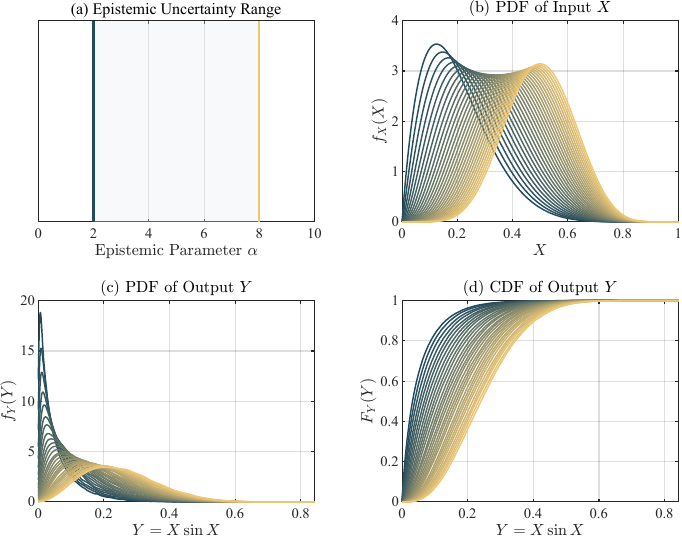}
	\caption{Visual representation of the hybrid uncertainty propagation process. (a) The epistemic parameter space $\Theta$. (b) The aleatory input space $\mathscr{P}_X$ represented by a family of conditional probability density functions (PDFs). (c) and (d) illustrate the output space $\mathscr{P}_Y$ through the corresponding propagated PDFs and cumulative distribution functions (CDFs), respectively, demonstrating the mapping from point-valued parameters to distribution-valued responses.}
	\label{fig:hybrid_mapping}
\end{figure}

In a hybrid uncertainty setting, the epistemic parameter $\theta$ fixes the aleatory input distribution $P_{X|\theta}$, which subsequently propagates through the system $Y=g(X)$ to form an output distribution $P_{Y|\theta}$, as visually illustrated in Figure \ref{fig:hybrid_mapping}:
\begin{align}
	\text{Epistemic Space:} \quad & \theta \in \Theta \subset \mathbb{R}^d \\
	\text{Aleatory Space:} \quad & \mathscr{P}_X = \{P_{X|\theta} \mid \theta \in \Theta\} \\
	\text{Output Space:} \quad & \mathscr{P}_Y = \{P_{Y|\theta} \mid \theta \in \Theta\}
\end{align}

This setting defines the updating problem addressed throughout this paper: given experimental observations of the output, infer the epistemic parameters $\theta$. Because the observable is a distribution induced by the aleatory inputs, the mapping from $\theta$ to the data distribution is many-to-one in the aleatory direction and the likelihood $p(\mathcal{D}_{\mathrm{obs}} \mid \theta)$ is analytically intractable for any non-trivial simulator $g$.

\subsection{Distance-based likelihood-free updating}
\label{sec:abc}

The established remedy for an intractable likelihood is Approximate Bayesian Computation (ABC) \cite{beaumont2002approximate, turner2012tutorial}. Given a prior $p(\theta)$ and a forward model, ABC replaces the likelihood with a distance-based similarity between simulated data $\mathbf{y}^*$ and the observation $\mathbf{y}_{\mathrm{obs}}$. A common formulation employs a Gaussian kernel,

\begin{equation}
	p_{\epsilon}(\theta|\mathbf y_{\mathrm{obs}})
	\propto
	\int
	\exp\!\left(
	-\frac{d(\mathbf y^*,\mathbf y_{\mathrm{obs}})^2}
	{\epsilon^2}
	\right)
	p(\mathbf y^*|\theta)
	p(\theta)
	\,d\mathbf y^*,
	\label{eq:abc_kernel}
\end{equation}
where $d(\cdot,\cdot)$ is a user-defined distance and $\epsilon>0$ controls the concentration of the posterior. Bayesian updating is thus entirely driven by the chosen distance measure, and samplers such as MCMC \cite{hastings1970monte}, TMCMC \cite{ching2007transitional, betz2016transitional} and TBQ \cite{song2026bayesian} can be employed to approximate the resulting posterior.

Two consequences follow directly from Eq.~\eqref{eq:abc_kernel}, and they mark the boundary of what this established method can deliver. First, its accuracy rests entirely on the quality of the distance $d$: for a distribution-valued output under hybrid uncertainty the distance must compare \emph{distributions}, not scalars. Second, every evaluation of the kernel requires simulating the model, so any noise in that simulation propagates directly into the likelihood---a point that becomes critical once a distributional distance is adopted, as discussed next.

\subsection{MMD as a distributional discrepancy and its existing use}
\label{sec:mmd}

Assume random vectors $\mathbf{x}, \mathbf{x}' \sim P$ and $\mathbf{y}, \mathbf{y}' \sim Q$ are independent copies drawn from two probability distributions $P$ and $Q$. The maximum mean discrepancy is defined as
\begin{equation}
	\mathrm{MMD}^2(P, Q) = \mathbb{E}_{\mathbf{x}, \mathbf{x}'} [k(\mathbf{x}, \mathbf{x}')] + \mathbb{E}_{\mathbf{y}, \mathbf{y}'} [k(\mathbf{y}, \mathbf{y}')] - 2\mathbb{E}_{\mathbf{x}, \mathbf{y}} [k(\mathbf{x}, \mathbf{y})],
	\label{MMD}
\end{equation}
where $\mathbb{E}_{\mathbf{x}, \mathbf{x}'}$ denotes the expectation with respect to the independent random variables $\mathbf{x}, \mathbf{x}' \sim P$, and $k(\cdot, \cdot)$ is a symmetric positive definite kernel function.

The effectiveness of MMD relies on characteristic kernels, under which the kernel mean embedding uniquely represents probability distributions in the associated reproducing kernel Hilbert space (RKHS) \cite{fukumizu2008characteristic}. In this case, MMD defines a proper metric on distributions, ensuring that any difference between them is detectable ($\mathrm{MMD}(P, Q) = 0$ if and only if $P = Q$) \cite{fukumizu2004dimensionality,fukumizu2007kernel}. Translation-invariant kernels are particularly convenient; notable examples are the Gaussian and Laplace kernels
\begin{align}
	k_\mathrm{G}(\mathbf{x}, \mathbf{x}'; \sigma) &= \exp \left( - \frac{\|\mathbf{x} - \mathbf{x}'\|^2}{2\sigma^2} \right), \label{eq:gaussian} \\
	k_\mathrm{L}(\mathbf{x}, \mathbf{x}'; \sigma) &= \exp \left( - \frac{\|\mathbf{x} - \mathbf{x}'\|}{\sigma} \right), \label{eq:laplace}
\end{align}
with the median heuristic \cite{gretton2005kernel} a common choice for the bandwidth $\sigma$. Given i.i.d. samples $\mathbf{X} = \{\mathbf{x}_i\}_{i=1}^n$ and $\mathbf{Y} = \{\mathbf{y}_j\}_{j=1}^m$ from $P$ and $Q$, an unbiased estimator is
\begin{equation}
	\widehat{\mathrm{MMD}}^2(\mathbf{X}, \mathbf{Y})
	= \frac{1}{n(n-1)} \sum_{i \neq i'} k(\mathbf{x}_i, \mathbf{x}_{i'})
	+ \frac{1}{m(m-1)} \sum_{j \neq j'} k(\mathbf{y}_j, \mathbf{y}_{j'})
	- \frac{2}{nm} \sum_{i,j} k(\mathbf{x}_i, \mathbf{y}_j).
	\label{uMMD}
\end{equation}

MMD is a principled candidate for the distance $d$ in Eq.~\eqref{eq:abc_kernel} when the output is distribution-valued, and it has been adopted for stochastic model updating under hybrid uncertainty \cite{kitahara2022nonparametric}: the surrogate likelihood is set to $\exp(-\widehat{\mathrm{MMD}}^2 / \epsilon^2)$, where the simulated side is an empirical output distribution obtained by drawing $n$ fresh random samples of the aleatory input at the candidate $\theta$. The difficulty lies in precisely that step. Because the aleatory inputs are random, each likelihood evaluation produces a \emph{different} empirical distribution, so the evaluated discrepancy is the sum of a systematic part, which depends on $\theta$, and a sampling part, which does not. When the sampling part is comparable to the systematic differences between competing parameters, the landscape of the surrogate likelihood over $\Theta$ is corrupted, and a noisy landscape necessarily degrades the posterior and any evidence derived from it. How large this corruption is in practice is quantified in the benchmark of Section~\ref{subsec:case1}, and it is the first motivation for the method developed next.

\subsection{PIT-based deterministic reformulation}
\label{sec:pit}

The two uncertainty types of Section \ref{sec:setting} can be separated through an auxiliary-variable reformulation introduced by Sankararaman and Mahadevan \cite{sankararaman2013separating} and since used chiefly for sensitivity analysis under hybrid uncertainty. For each uncertain input component one introduces an auxiliary variable $\xi \sim \mathcal{U}(0,1)$ through
\begin{equation}
	\xi = F_X(X \mid \theta), \qquad X = F_X^{-1}(\xi \mid \theta),
	\label{eq:pit}
\end{equation}
where $F_X(\cdot \mid \theta)$ denotes the conditional CDF of the input and $F_X^{-1}(\cdot \mid \theta)$ its generalized inverse, i.e. the quantile function; for correlated inputs the same construction applies through a Rosenblatt--Nataf transform, so that $\xi$ is uniform over the unit hypercube. The auxiliary variable $\xi$ is statistically independent of $\theta$ and carries the entire aleatory variability in a distribution-free form, while all epistemic uncertainty is isolated in $\theta$. Substituting Eq.~\eqref{eq:pit} into the response model yields
\begin{equation}
	Y = g\!\left(F_X^{-1}(\xi \mid \theta)\right) = h(\xi, \theta),
	\label{eq:deterministic}
\end{equation}
a map that is deterministic in the joint variable $(\xi, \theta)$. This reformulation is adopted in this paper as the representation layer on which the proposed updating method is built; on its own it does not resolve the resampling noise of discrepancy evaluation, because existing uses of the PIT reformulation still draw $\xi$ afresh at every evaluation.

\subsection{Limitations of existing approaches}
\label{sec:limits}

The review above can be condensed into two limitations that any satisfactory method must resolve.

{(P1) Intractable likelihood for distribution-valued outputs.} Under hybrid uncertainty the observable is a distribution, not a scalar; neither the likelihood nor its kernel-form approximations in Eq.~\eqref{eq:abc_kernel} can be written in closed form, and scalar distances cannot compare distributions.

{(P2) Sampling noise in the discrepancy.} Existing MMD-based updating redraws the simulator at every likelihood evaluation. The resulting sampling fluctuation is not a second-order nuisance: it is a substantial fraction of the parameter-discriminating signal (quantified in Section~\ref{subsec:case1}), so the landscape is corrupted at the source, necessarily degrading the posterior and any evidence derived from it.

The method developed next is constructed so that each of its components answers one of these limitations.

\section{PIT-MMD Updating via TMCMC}
\label{sec:method}

\subsection{Deterministic Discrepancy Formulation via Frozen Quantile Particles and MMD}
\label{sec:particles}

The reformulation of Eq.~\eqref{eq:deterministic} becomes computationally advantageous once the latent variable is discretized. Instead of drawing fresh random samples of $X$ at every model evaluation---the step that creates the noise in (P2)---we fix a set of $M$ stratified quantile particles $\{\xi_i\}_{i=1}^{M}$ covering the unit hypercube, generated once by Latin hypercube sampling and then frozen. For any candidate $\theta$, propagating the frozen particles through Eq.~\eqref{eq:deterministic} yields a particle cloud $\mathcal{C}_\theta = \{h(\xi_i, \theta)\}_{i=1}^{M}$ whose empirical distribution approximates $P_{Y|\theta}$. Because the particles are never re-drawn, $\mathcal{C}_\theta$ is a deterministic function of $\theta$: identical parameters yield identical clouds, and nearby parameters yield nearby clouds. As quantified in Section~\ref{subsec:case1}, this common-random-number construction removes the sampling fluctuation that would otherwise contaminate every likelihood evaluation, thereby eliminating this resampling noise by construction rather than compensating for it afterwards.

\begin{figure}[htbp]
	\centering
	\includegraphics{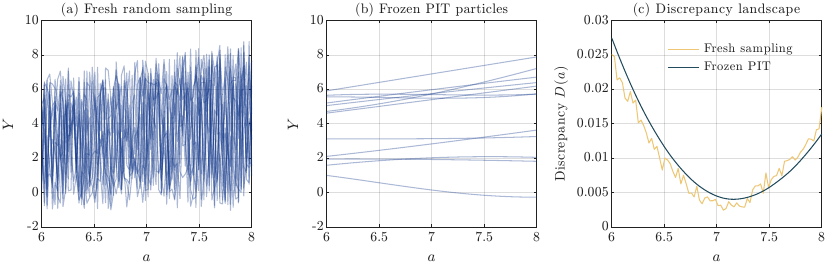}
	\caption{Mechanism of the frozen-particle construction on the 2D synthetic model (one-dimensional slice in $a$ at fixed $b = b^{*}$). (a) Under fresh random sampling the tracked particles fluctuate across $\theta$, producing a stochastic cloud; (b) under frozen PIT particles the cloud varies deterministically with $\theta$; (c) the resulting discrepancy $D(a)$ is smooth and cleanly localized at the minimum, in contrast to the noisy fresh-sampling curve.}
	\label{fig:mechanism}
\end{figure}

With the noise removed, the discrepancy itself must compare the particle cloud $\mathcal{C}_\theta$ with the experimental samples $\mathcal{D}_{\mathrm{obs}} = \{\mathbf{y}^{\mathrm{obs}}_j\}_{j=1}^{N}$, which is where the MMD of Eq.~\eqref{MMD} enters: $P$ is represented by the empirical distribution of $\mathcal{C}_\theta$ and $Q$ by that of $\mathcal{D}_{\mathrm{obs}}$. A single fixed kernel bandwidth is rarely adequate when the response components differ in scale, so we adopt a multi-scale Gaussian kernel
\begin{equation}
	\bar{k}(\mathbf{y}, \mathbf{y}') = \frac{1}{K} \sum_{k=1}^{K} \exp\!\left( - \frac{\|\mathbf{y} - \mathbf{y}'\|^2}{2 \sigma_k^2} \right),
	\label{eq:mskernel}
\end{equation}
with bandwidths $\sigma_k$ anchored at the median heuristic $\sigma_{\mathrm{med}} = \sqrt{\mathrm{median}\,\{\|\mathbf{y}^{\mathrm{obs}}_j - \mathbf{y}^{\mathrm{obs}}_{j'}\|^2\}}$ of the observed pairwise distances \cite{gretton2005kernel} and scaled by $\{0.25, 0.5, 1, 2\}$, so that both coarse and fine structure of the distributions contribute to the discrepancy. The resulting point estimate of the squared MMD between the two empirical distributions is
\begin{equation}
	D(\theta) = \frac{1}{M^2} \sum_{i,j} \bar{k}\!\left(\mathbf{y}_i(\theta), \mathbf{y}_j(\theta)\right) + \frac{1}{N^2} \sum_{j,j'} \bar{k}\!\left(\mathbf{y}^{\mathrm{obs}}_j, \mathbf{y}^{\mathrm{obs}}_{j'}\right) - \frac{2}{MN} \sum_{i,j} \bar{k}\!\left(\mathbf{y}_i(\theta), \mathbf{y}^{\mathrm{obs}}_j\right),
	\label{eq:disc}
\end{equation}
where $\mathbf{y}_i(\theta) = h(\xi_i, \theta)$. The observation-dependent second term is a constant precomputed once, and the evaluation cost is $\mathcal{O}(M^2 + MN)$ kernel evaluations, which is negligible for the particle sizes used here ($M = 200$); no low-rank feature approximation is required at this scale.

\begin{figure}[htbp]
	\centering
	\includegraphics{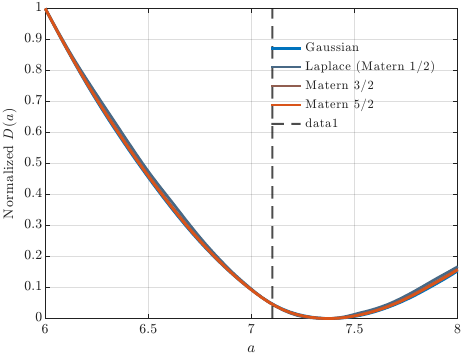}
	\caption{Kernel ablation on the 2D benchmark: normalized discrepancy landscape $D(a)$ for the Gaussian, Laplace (Mat\'ern-$1/2$), Mat\'ern-$3/2$ and Mat\'ern-$5/2$ kernels, all with median-heuristic multi-scale bandwidths. The Gaussian kernel yields the deepest, best-localized minimum at the true value $a^{*}$.}
	\label{fig:kernel_comparison}
\end{figure}

The Gaussian kernel is chosen among common alternatives---the Laplace kernel (which coincides with a Mat\'ern-$1/2$ kernel) and the Mat\'ern-$3/2$ and Mat\'ern-$5/2$ kernels---through a controlled ablation on the benchmark of Section \ref{subsec:case1}. With the same median-heuristic multi-scale bandwidths, the Gaussian kernel yields the largest discrimination ratio between the true parameter and the prior discrepancy distribution ($2.55$, versus $2.45$ for Mat\'ern-$5/2$, $2.37$ for Mat\'ern-$3/2$ and $2.09$ for Laplace), as illustrated in Figure \ref{fig:kernel_comparison}.

Replacing the generic distance in the ABC kernel of Eq.~\eqref{eq:abc_kernel} by the multi-scale MMD of Eq.~\eqref{eq:disc} yields the energy-form surrogate likelihood
\begin{equation}
	L(\mathcal{D}_{\mathrm{obs}} \mid \theta) = \exp\!\left( - \frac{D(\theta)}{2 \tau^2} \right),
	\qquad
	p_{\tau}(\theta \mid \mathcal{D}_{\mathrm{obs}}) \propto L(\mathcal{D}_{\mathrm{obs}} \mid \theta)\, p(\theta),
	\label{eq:abc_surrogate}
\end{equation}
where $\tau > 0$ plays the role of the ABC bandwidth and controls the concentration of the approximate posterior. Because $D(\theta)$ is deterministic under the PIT construction, $L$ is a smooth function of $\theta$, and $\tau$ is set by a data-driven heuristic calibration rule: we evaluate $D(\theta)$ at a batch of prior draws and set $\tau^2$ to one sixth of the range between its $90\%$ quantile and its minimum, so that the likelihood spans a few nats across the prior domain, enough to discriminate without over-concentrating. This calibration is reported for each case study in Section \ref{sec:case}.

\subsection{Posterior inference via TMCMC}
\label{sec:inference}

The posterior of Eq.~\eqref{eq:abc_surrogate} is sampled with Transitional Markov Chain Monte Carlo \cite{ching2007transitional}, which constructs a sequence of intermediate distributions $\pi_j(\theta) \propto L^{\beta_j} p(\theta)$ bridging the prior ($\beta_0 = 0$) to the posterior ($\beta_n = 1$). We employ a fixed quadratic annealing schedule $\beta_j = (j/n)^2$, systematic resampling, and per-stage Metropolis moves whose proposal covariance is adapted from the weighted particle covariance. TMCMC additionally returns an estimate of the log evidence of the surrogate model,
\begin{equation}
	\log Z = \sum_{j=0}^{n-1} \log \left( \frac{1}{N_p} \sum_{i=1}^{N_p} \exp\!\left( (\beta_{j+1} - \beta_j)\, \ell(\theta_i^{(j)}) \right) \right),
	\label{eq:evidence}
\end{equation}
where $\ell = \log L$ and $\theta_i^{(j)}$ are the particles of stage $j$. This closes the loop opened in (P3): the same deterministic likelihood that serves the posterior also yields the evidence at no additional cost. Algorithm \ref{alg:modular_abc} summarizes the complete procedure.

\begin{algorithm}[htbp]
	\caption{PIT-MMD Bayesian Inference via TMCMC}
	\label{alg:modular_abc}
	\begin{algorithmic}[1]
		\Require Prior $p(\theta)$, simulator $g$, frozen quantile particles $\{\xi_i\}_{i=1}^M$, observations $\mathcal{D}_{\mathrm{obs}}$, bandwidths $\{\sigma_k\}$, scale $\tau$, stages $n$.
		\Ensure Posterior samples and log evidence.
		\State Precompute $C_{\mathrm{obs}} = N^{-2} \sum_{j,j'} \bar{k}(\mathbf{y}^{\mathrm{obs}}_j, \mathbf{y}^{\mathrm{obs}}_{j'})$.
		\State Draw $\theta_i^{(0)} \sim p(\theta)$, $i = 1, \dots, N_p$; evaluate $\ell(\theta_i^{(0)})$ via Eqs.~\eqref{eq:disc} and \eqref{eq:abc_surrogate}.
		\For{$j = 0, \dots, n-1$}
		\State Reweight particles with exponents $(\beta_{j+1} - \beta_j)\, \ell$; accumulate $\log Z$ (Eq.~\eqref{eq:evidence}).
		\State Resample systematically; compute weighted covariance $\Sigma_j$ and set $\mathcal{L}_j = c\,\mathrm{chol}(\Sigma_j)$.
		\For{$i = 1, \dots, N_p$}
		\State Propose $\theta' = \theta_i + \mathcal{L}_j \, \mathbf{z}$, $\mathbf{z} \sim \mathcal{N}(0, I)$; evaluate $\ell(\theta')$.
		\State Accept with probability $\min\!\left(1, \exp\!\left( \beta_{j+1} \left[ \ell(\theta') - \ell(\theta_i) \right] \right) \right)$.
		\EndFor
		\EndFor
		\State \Return posterior particles $\{\theta_i^{(n)}\}$ and $\log Z$.
	\end{algorithmic}
\end{algorithm}

\section{Application cases}
\label{sec:case}

The three case studies are arranged to verify the causal claims of the proposed method in increasing order of realism. Case 1 verifies the central claim that the frozen-particle construction removes the discrepancy noise and thereby improves the inference, by controlled comparison against the random-sampling control on a synthetic model with a known truth. Case 2 verifies scalability to functional outputs. Case 3 verifies applicability to a real engineering benchmark with an official black-box simulator and experimental data.

\subsection{Case 1: Controlled validation on a 2D synthetic model}
\label{subsec:case1}

We first consider a canonical two-dimensional synthetic model to isolate the effect of the frozen-particle construction. The system contains two epistemic parameters, $\theta = [a, b]^\top$, and a three-dimensional aleatory input $X = [X_1, X_2, X_3]^\top$ whose distributions are governed by $\theta$:
\begin{equation}
	X_1 \sim \mathcal{U}(-a, a), \quad X_2 \sim \mathcal{U}(-\pi+b, \pi+b), \quad X_3 \sim \mathcal{U}(-ab, ab),
\end{equation}
and the output is the highly non-linear mapping $Y = [Y_1, Y_2, Y_3]^\top$:
\begin{equation}
	\begin{cases}
		Y_1 = \sin(X_1) + a \sin^2(X_2) + b X_3^4 \sin(X_1) \\
		Y_2 = \cos(a X_1) - b X_2 \\
		Y_3 = X_1^2 + X_2^2 + X_3^2
	\end{cases}
\end{equation}
The prior is $a \in [6.0, 8.0]$, $b \in [0.05, 0.15]$, and the true parameter, drawn uniformly from the prior with a fixed random seed, is $\theta^{*} = [7.1016, 0.1208]^\top$. The PIT map of Eq.~\eqref{eq:pit} acts component-wise through the inverse CDFs of the three uniform inputs, so each particle is a triple $\xi_i \in [0,1]^3$. We use $M = 200$ frozen particles and $N = 50$ observed output samples generated at $\theta^{*}$; the multi-scale kernel bandwidths are anchored at the median heuristic of the observed cloud.

\begin{figure}[htbp]
	\centering
	\includegraphics{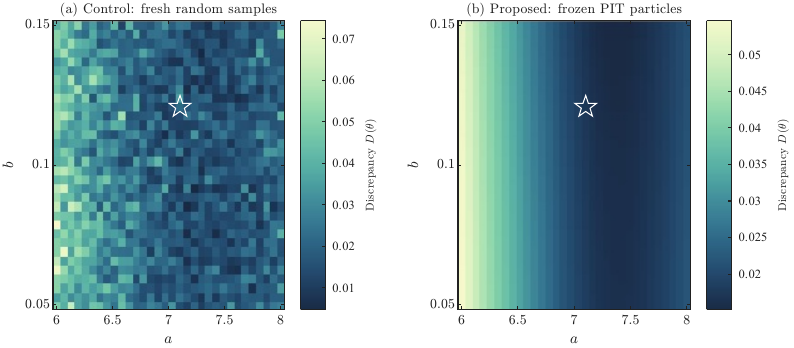}
	\caption{Discrepancy landscape $D(\theta)$ for the 2D synthetic model over the prior domain. (a) Control scheme with fresh random sampling at every evaluation: the landscape is corrupted by sampling noise. (b) Proposed PIT construction with frozen quantile particles: the landscape is deterministic and smooth, and its minimum is cleanly localized at the true parameter $\theta^{*}$ (orange pentagram). The overlaid white points are posterior samples.}
	\label{fig:surface}
\end{figure}

Figure \ref{fig:surface} compares the discrepancy landscape of the two schemes on a regular grid of the prior domain. Under the control scheme, which draws a fresh random sample of $X$ at every evaluation, the same $\theta$ yields substantially different discrepancies from one evaluation to the next: the within-point standard deviation over repeated evaluations is $0.0067$, i.e., about $29\%$ of the mean discrepancy level. This fluctuation is a substantial fraction of the depth of the discrepancy well around $\theta^{*}$, and it renders the landscape noisy. Under the PIT construction the discrepancy is exactly deterministic, the landscape is smooth, and the minimum is cleanly localized around the true parameter. The TMCMC posterior samples overlaid on the PIT landscape concentrate in this low-discrepancy region.

\begin{figure}[htbp]
	\centering
	\includegraphics{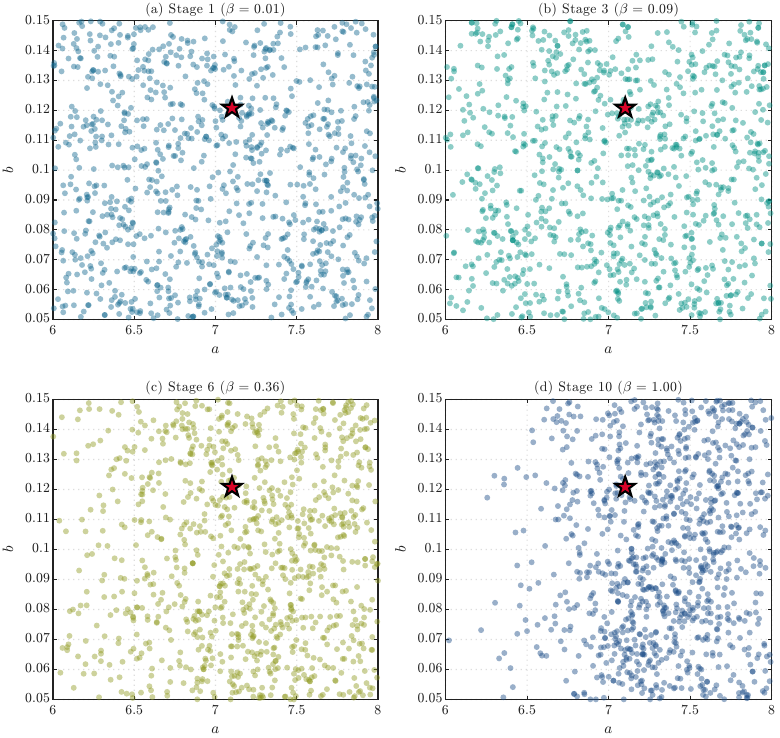}
	\caption{TMCMC inference for the 2D synthetic model. (a) Evolution of the particle population across the $n = 10$ annealing stages, colored by the annealing parameter $\beta$. (b) Joint posterior samples; the orange pentagram marks the true parameter $\theta^{*}$.}
	\label{fig:trace}
\end{figure}

The posterior is then sampled by TMCMC with $N_p = 1000$ particles and $n = 10$ annealing stages, which completes in $8.4$ s and returns a log evidence of $-2.40$. Figure \ref{fig:trace} shows the annealing evolution of the particle population and the resulting joint posterior. The marginal posterior mean is $[7.32, 0.0998]^\top$ with standard deviations $[0.37, 0.029]$, enclosing the true value $[7.10, 0.121]^\top$ within one posterior standard deviation in both coordinates; the wider spread of $b$ reflects its weaker influence on the model output.

\subsection{Case 2: High-dimensional transient damped harmonic oscillator}
\label{subsec:case2}

\begin{figure}[htbp]
	\centering
	\includegraphics{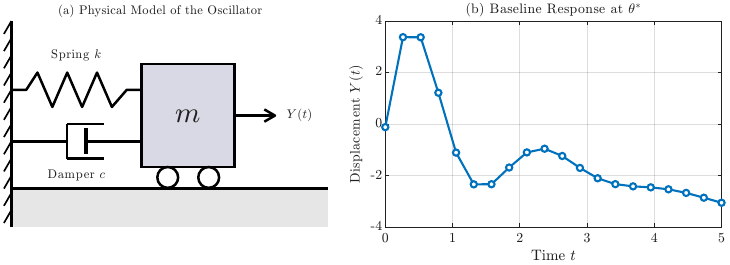}
	\caption{(a) Schematic diagram of the transient mass-spring-damper oscillator. (b) The deterministic baseline response evaluated at the true parameter $\theta^{*}$.}
	\label{fig:oscillator_baseline}
\end{figure}

To verify scalability to functional outputs, we investigate a 12-dimensional hybrid uncertainty problem based on a transient mass-spring-damper oscillator. The displacement response $Y(t)$ is governed by 5 aleatory variables ($X_1 \dots X_5$) and 7 epistemic parameters ($\theta_1 \dots \theta_7$):
\begin{equation}
	Y(t) = 10 X_1 e^{-2 X_2 t} \sin(2\pi X_3 t) + X_4 \frac{t}{5} + X_5, \quad t \in [0, 5],
\end{equation}
where the output is a 20-dimensional time-series vector evaluated at uniformly spaced intervals. The physical model and the deterministic baseline response evaluated at the true parameter are illustrated in Figure \ref{fig:oscillator_baseline}.

The epistemic vector $\theta = [\mu_1, \sigma^2_1, \mu_4, \sigma^2_4, \mu_5, \sigma^2_5, \rho]^\top$ controls the hyper-parameters of the underlying aleatory distributions: $X_1$ follows a Beta distribution parameterized by $\mu_1$ and $\sigma^2_1$, $X_2$ and $X_3$ are standard uniform, and $X_4$ and $X_5$ follow a correlated bivariate Gaussian with correlation coefficient $\rho$. The PIT map of Eq.~\eqref{eq:pit} acts through the inverse Beta CDF, the two uniform CDFs, and the bivariate normal quantiles, so each frozen particle is a five-dimensional uniform vector. The prior boundaries and the true parameter values (drawn uniformly with a fixed random seed) are: $\theta^{*} = [0.644,\ 0.0374,\ -2.933,\ 3.675,\ -0.116,\ 2.448,\ 0.479]^\top$ over the prior $\mu_1 \in [0.60, 0.80]$, $\sigma^2_1 \in [0.02, 0.04]$, $\mu_4, \mu_5 \in [-5, 5]$, $\sigma^2_4, \sigma^2_5 \in [0.0025, 4]$, and $\rho \in [-0.9, 0.9]$. We use $M = 200$ frozen particles and $N = 50$ observed 20-dimensional time-series samples generated at $\theta^{*}$.

\begin{figure}[htbp]
	\centering
	\includegraphics{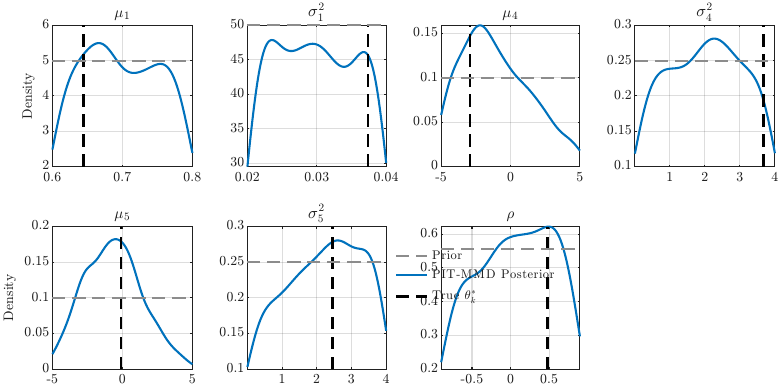}
	\caption{Marginal posterior distributions for the 7 epistemic parameters of the damped harmonic oscillator. The uniform priors (dashed black) are updated by PIT-MMD (blue), and the dashed orange lines mark the true parameter values.}
	\label{fig:oscillator_marginals}
\end{figure}

The discrepancy at the true parameter ($D = 0.0093$) is separated from the $90\%$ quantile of the prior discrepancy landscape ($D = 0.518$) by more than a factor of fifty, so the landscape is strongly informative. TMCMC with $N_p = 1000$ particles and $n = 10$ stages completes in $9.8$ s ($\log Z = -0.99$). Figure \ref{fig:oscillator_marginals} shows the marginal posterior of all 7 parameters: the mean parameter $\mu_1$ and the variance $\sigma^2_1$ are sharply identified, while $\sigma^2_4$, $\sigma^2_5$, $\rho$ and the means $\mu_4$, $\mu_5$ remain wider, reflecting the weaker sensitivity of the transient response to these hyper-parameters.

\subsection{Case 3: NASA Langley UQ Challenge, Subproblem A}
\label{subsec:case3}

Finally, the framework is applied to Subproblem A of the NASA Langley Multidisciplinary Uncertainty Quantification Challenge, a real engineering benchmark in which a scalar intermediate response of a launch-vehicle system, $x_1 = h_1(p_1, \dots, p_5)$, is evaluated through an official black-box simulator. The five uncertain inputs belong to the three challenge categories: $p_1$ is an aleatory Beta variable whose mean and variance are epistemic ($E[p_1] \in [0.6, 0.8]$, $V[p_1] \in [0.02, 0.04]$), $p_2$ is a fixed but unknown epistemic constant ($p_2 \in [0, 1]$), $p_3$ is a fully prescribed uniform variable, and $p_4$, $p_5$ form a correlated bivariate Gaussian whose means, variances and correlation are epistemic ($E[p_i] \in [-5, 5]$, $V[p_i] \in [0.0025, 4]$, $|\rho| \le 1$). The complete epistemic vector is therefore 8-dimensional, $\theta = [E[p_1], V[p_1], p_2, E[p_4], V[p_4], E[p_5], V[p_5], \rho]^\top$, and every component except $p_3$ is directly coupled to the aleatory description---precisely the hybrid structure addressed by the proposed framework.

The challenge provides $n = 25$ observations of $x_1$ generated from the true uncertainty model for task A1 and another $n = 25$ for validation in task A2; following task A3, we merge both sets into a single observation record of $N = 50$ scalar samples (sample mean $0.232$, standard deviation $0.147$). Each frozen particle $\xi_i \in [0,1]^4$ encodes the Beta quantile of $p_1$, the uniform quantile of $p_3$, and the two standard-normal quantiles driving the correlated pair $(p_4, p_5)$; the epistemic constant $p_2 = \theta_3$ is common to all particles. With $M = 200$ particles, the discrepancy of Eq.~\eqref{eq:disc} is evaluated through the official simulator, whose vectorized evaluation of an entire particle cloud requires only a few milliseconds after initialization.

\begin{figure}[htbp]
	\centering
	\includegraphics{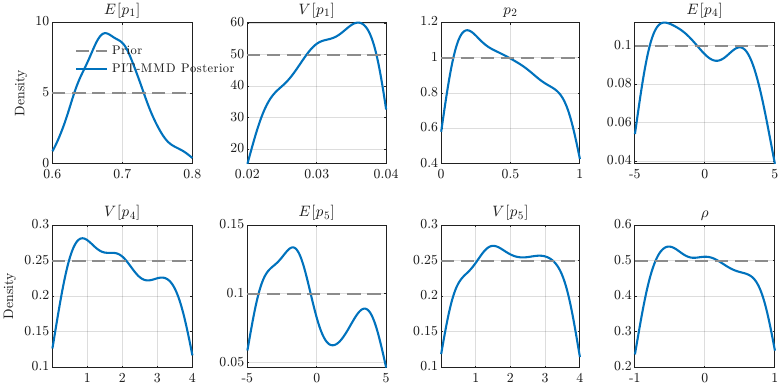}
	\caption{Posterior marginals of the 8 epistemic parameters of the NASA LaRC challenge (Subproblem A). The uniform priors (dashed black) are updated by PIT-MMD (blue) from the $N = 50$ merged observations. The parameters that drive the observable ($E[p_1]$, $V[p_1]$) are sharply identified, while weakly informed directions remain prior-like.}
	\label{fig:nasa_marginals}
\end{figure}

\begin{figure}[htbp]
	\centering
	\includegraphics{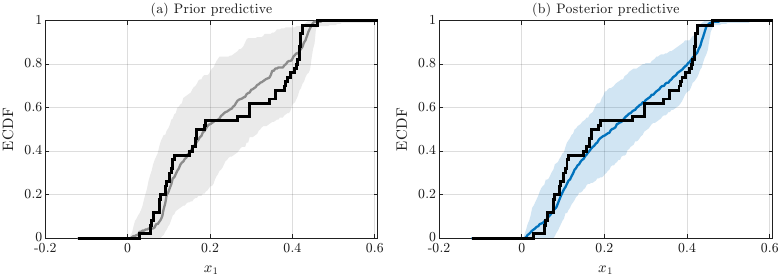}
	\caption{Predictive check for the NASA LaRC challenge in the observation space. Empirical CDF of the $N = 50$ observed samples of $x_1$ (black stairs), together with the median (line) and the 2.5\%--97.5\% band (shaded) of the predictive ECDFs under (a) the prior and (b) the posterior.}
	\label{fig:nasa_predictive}
\end{figure}

Figure \ref{fig:nasa_marginals} shows the resulting posterior marginals obtained with $N_p = 1500$ particles and $n = 10$ stages ($33.1$ s, $\log Z = -1.49$). The data deliver a genuine epistemic reduction for the parameters that drive the observable: the posterior of $E[p_1]$ concentrates at $0.68 \pm 0.04$ (a $31\%$ reduction of the prior standard deviation) and $V[p_1]$ at $0.031 \pm 0.005$, while the variances $V[p_4]$ and $V[p_5]$ are also updated. By contrast, $p_2$, $E[p_4]$, $E[p_5]$ and $\rho$ remain close to their priors: fifty scalar observations of a single intermediate response simply do not carry enough information to identify these directions, and the proposed method reports this honestly through wide, prior-like marginals instead of artificial concentration. Figure \ref{fig:nasa_predictive} confirms the calibration in the observation space: the posterior predictive ECDF of $x_1$ collapses onto the observed ECDF, in sharp contrast to the wide prior predictive band.

\subsection{Computational cost}
\label{sec:cost}

All experiments were run on a desktop workstation with an AMD Ryzen 7 5800H CPU (8 cores, 16 threads) and 16 GB of memory, using a single-threaded MATLAB implementation and the same TMCMC configuration ($n = 10$ annealing stages) throughout. Table \ref{tab:cost} summarizes the computational cost of the three case studies.

\begin{table}[htbp]
	\centering
	\caption{Computational cost of the PIT-MMD inference for the three case studies.}
	\label{tab:cost}
	\begin{tabular}{lcccccc}
		\toprule
		\textbf{Case} & $d$ & $M$ & $N$ & $N_p$ & \textbf{Stages} & \textbf{TMCMC time} \\
		\midrule
		2D synthetic & $2$ & $200$ & $50$ & $1000$ & $10$ & $8.4$ s \\
		Oscillator & $7$ & $200$ & $50$ & $1000$ & $10$ & $9.8$ s \\
		NASA LaRC & $8$ & $200$ & $50$ & $1500$ & $10$ & $33.1$ s \\
		\bottomrule
	\end{tabular}
\end{table}

where $d$ is the number of epistemic parameters, $M$ the number of frozen particles, $N$ the number of observations, and $N_p$ the TMCMC population size. Each discrepancy evaluation costs $\mathcal{O}(M^2 + MN)$ kernel evaluations; with $M = 200$ and $N = 50$ this is about $5 \times 10^4$ operations, and the memory footprint of the entire particle system is below $0.1$ MB in all cases. For the NASA benchmark the dominant cost is the call to the official black-box simulator, which evaluates a full particle cloud in a single vectorized call in a few milliseconds after initialization.

Table \ref{tab:compare} compares the frozen-particle (proposed) construction against the fresh-resampling (traditional) scheme on the 2D benchmark, with all other settings identical. The two schemes have essentially identical time and memory; the difference lies in the effect: the frozen construction removes the sampling noise and thereby yields a narrower posterior ($\mathrm{std}(a) = 0.370$ versus $0.422$) on the same deterministic landscape. A fresh-resampling scheme that redraws stratified (Latin hypercube) samples at every evaluation would instead pay the stratification cost at each step, running about four times slower ($35.7$ s versus $8.6$ s).

\begin{table}[htbp]
	\centering
	\caption{Frozen-particle (proposed) versus fresh-resampling (traditional) on the 2D benchmark, with identical TMCMC settings.}
	\label{tab:compare}
	\begin{tabular}{lcccc}
		\toprule
		\textbf{Scheme} & \textbf{Time} & \textbf{Memory} & $\mathrm{std}(a)$ & \textbf{Noise} \\
		\midrule
		Fresh resampling (traditional) & $8.6$ s & $<0.1$ MB & $0.422$ & $\approx 29\%$ \\
		Frozen PIT (proposed) & $8.6$ s & $<0.1$ MB & $0.370$ & $0$ \\
		\bottomrule
	\end{tabular}
\end{table}

\section{Conclusion}
\label{sec:conclusion}

Following the line of existing methods, their limitations, the proposed remedy, and its validation, this paper identified sampling noise as the structural weakness of discrepancy-based likelihood-free updating under hybrid uncertainty: existing MMD-based schemes redraw the simulator at every likelihood evaluation, and the resulting fluctuation corrupts the landscape from which the posterior and the evidence are computed. The proposed PIT-MMD framework eliminates this resampling noise by construction, by reformulating the simulator through the probability integral transform and freezing a set of stratified quantile particles, so that the simulated output distribution, the multi-scale MMD discrepancy, and the energy-form surrogate likelihood all become deterministic functions of the epistemic parameters. TMCMC then delivers both the posterior and the surrogate-model evidence on this smooth landscape. The framework was validated on a two-dimensional benchmark with a controlled comparison against random sampling, on a high-dimensional transient oscillator, and on Subproblem A of the NASA Langley Multidisciplinary Uncertainty Quantification Challenge with the official black-box simulator and experimental data. The complete Bayesian analysis of the real benchmark, including evidence estimation, completed in about half a minute on a desktop workstation.

Future work will focus on two directions. First, adaptive and Bayesian calibration of the discrepancy scale $\tau$ will be investigated, together with the effect of the number of observations on the identifiability of the epistemic parameters. Second, the deterministic PIT landscape opens the door to gradient-based and surrogate-accelerated exploration of the epistemic space, enabling application to larger-scale engineering models.

	\bibliographystyle{unsrt}
	\bibliography{references}

@article{hastings1970monte,
    author = {Hastings, W. K.},
    title = {Monte Carlo sampling methods using Markov chains and their applications},
    journal = {Biometrika},
    volume = {57},
    number = {1},
    pages = {97-109},
    year = {1970},
    month = {04},
    issn = {0006-3444},
    doi = {10.1093/biomet/57.1.97},
    url = {https://doi.org/10.1093/biomet/57.1.97},
    eprint = {https://academic.oup.com/biomet/article-pdf/57/1/97/23940249/57-1-97.pdf},
}

@article{turner2012tutorial,
  title={A tutorial on approximate Bayesian computation},
  author={Turner, Brandon M and Van Zandt, Trisha},
  journal={Journal of Mathematical Psychology},
  volume={56},
  number={2},
  pages={69--85},
  year={2012},
  publisher={Elsevier}
}

@article{beaumont2002approximate,
  title={Approximate Bayesian computation in population genetics},
  author={Beaumont, Mark A and Zhang, Wenyang and Balding, David J},
  journal={Genetics},
  volume={162},
  number={4},
  pages={2025--2035},
  year={2002},
  publisher={Oxford University Press}
}

@article{fukumizu2008characteristic,
	title={Characteristic kernels on groups and semigroups},
	author={Fukumizu, Kenji and Gretton, Arthur and Sch{\"o}lkopf, Bernhard and Sriperumbudur, Bharath K},
	journal={Advances in neural information processing systems},
	volume={21},
	year={2008}
}

@article{fukumizu2004dimensionality,
	title={Dimensionality reduction for supervised learning with reproducing kernel Hilbert spaces},
	author={Fukumizu, Kenji and Bach, Francis R and Jordan, Michael I},
	journal={Journal of Machine Learning Research},
	volume={5},
	number={Jan},
	pages={73--99},
	year={2004}
}

@article{fukumizu2007kernel,
	title={Kernel measures of conditional dependence},
	author={Fukumizu, Kenji and Gretton, Arthur and Sun, Xiaohai and Sch{\"o}lkopf, Bernhard},
	journal={Advances in neural information processing systems},
	volume={20},
	year={2007}
}

@article{gretton2005kernel,
	title={Kernel methods for measuring independence},
	author={Gretton, Arthur and Herbrich, Ralf and Smola, Alexander and Bousquet, Olivier and Sch{\"o}lkopf, Bernhard},
	year={2005},
	publisher={MIT Press}
}

@article{xu2022importance,
  title={Importance Weighting Approach in Kernel Bayes' Rule},
  author={Xu, Liyuan and Chen, Yutian and Doucet, Arnaud and Gretton, Arthur},
  journal={arXiv preprint arXiv:2202.02474},
  year={2022}
}

@article{ching2007transitional,
  title={Transitional Markov chain Monte Carlo method for Bayesian model updating, model class selection, and model averaging},
  author={Ching, Jianye and Chen, Yi-Chu},
  journal={Journal of engineering mechanics},
  volume={133},
  number={7},
  pages={816--832},
  year={2007},
  publisher={American Society of Civil Engineers}
}

@article{betz2016transitional,
  title={Transitional Markov chain Monte Carlo: observations and improvements},
  author={Betz, Wolfgang and Papaioannou, Iason and Straub, Daniel},
  journal={Journal of Engineering Mechanics},
  volume={142},
  number={5},
  pages={04016016},
  year={2016},
  publisher={American Society of Civil Engineers}
}

@article{song2026bayesian,
  title={Bayesian active learning for Bayesian model updating: The art of acquisition functions and beyond},
  author={Song, Jingwen and Wei, Pengfei},
  journal={Mechanical Systems and Signal Processing},
  volume={251},
  pages={114237},
  year={2026},
  publisher={Elsevier}
}

@article{gretton2012kernel,
  title={A kernel two-sample test},
  author={Gretton, Arthur and Borgwardt, Karsten M and Rasch, Malte J and Sch{\"o}lkopf, Bernhard and Smola, Alexander},
  journal={The journal of machine learning research},
  volume={13},
  number={1},
  pages={723--773},
  year={2012},
  publisher={JMLR. org}
}

@article{kitahara2022nonparametric,
  title={Nonparametric Bayesian stochastic model updating with hybrid uncertainties},
  author={Kitahara, Masaru and Bi, Sifeng and Broggi, Matteo and Beer, Michael},
  journal={Mechanical Systems and Signal Processing},
  volume={163},
  pages={108195},
  year={2022},
  publisher={Elsevier}
}

@article{bi2023stochastic,
  title={Stochastic model updating with uncertainty quantification: an overview and tutorial},
  author={Bi, Sifeng and Beer, Michael and Cogan, Scott and Mottershead, John},
  journal={Mechanical Systems and Signal Processing},
  volume={204},
  pages={110784},
  year={2023},
  publisher={Elsevier}
}

@article{oberkampf2004challenge,
  title={Challenge problems: uncertainty in system response given uncertain parameters},
  author={Oberkampf, William L and Helton, Jon C and Joslyn, Cliff A and Wojtkiewicz, Steven F and Ferson, Scott},
  journal={Reliability Engineering \& System Safety},
  volume={85},
  number={1-3},
  pages={11--19},
  year={2004},
  publisher={Elsevier}
}

@article{roy2011comprehensive,
  title={A comprehensive framework for verification, validation, and uncertainty quantification in scientific computing},
  author={Roy, Christopher J and Oberkampf, William L},
  journal={Computer methods in applied mechanics and engineering},
  volume={200},
  number={25-28},
  pages={2131--2144},
  year={2011},
  publisher={Elsevier}
}

@inproceedings{zarei2023indoor,
  title={Indoor UAV object detection algorithms on three processors: implementation test and comparison},
  author={Zarei, Mohammad and Moshayedi, Ata Jahangir and Zhong, Yangwan and Khan, Amir Sohail and Kolahdooz, Amin and Andani, Mehran Emadi},
  booktitle={2023 3rd international conference on consumer electronics and computer engineering (ICCECE)},
  pages={812--819},
  year={2023},
  organization={IEEE}
}

@book{moore2009introduction,
  title={Introduction to interval analysis},
  author={Moore, Ramon E and Kearfott, R Baker and Cloud, Michael J},
  year={2009},
  publisher={SIAM}
}

@article{faes2020recent,
  title={Recent Trends in the Modeling and Quantification of Non-probabilistic Uncertainty: M. Faes, D. Moens},
  author={Faes, Matthias and Moens, David},
  journal={Archives of Computational Methods in Engineering},
  volume={27},
  number={3},
  pages={633--671},
  year={2020},
  publisher={Springer}
}

@article{calvi2005uncertainty,
  title={Uncertainty-based loads analysis for spacecraft: Finite element model validation and dynamic responses},
  author={Calvi, Adriano},
  journal={Computers \& structures},
  volume={83},
  number={14},
  pages={1103--1112},
  year={2005},
  publisher={Elsevier}
}

@article{beaumont2009adaptive,
  title={Adaptive approximate Bayesian computation},
  author={Beaumont, Mark A and Cornuet, Jean-Marie and Marin, Jean-Michel and Robert, Christian P},
  journal={Biometrika},
  volume={96},
  number={4},
  pages={983--990},
  year={2009},
  publisher={Oxford University Press}
}

@article{nakagome2013kernel,
  title={Kernel approximate Bayesian computation in population genetic inferences},
  author={Nakagome, Shigeki and Fukumizu, Kenji and Mano, Shuhei},
  journal={Statistical applications in genetics and molecular biology},
  volume={12},
  number={6},
  pages={667--678},
  year={2013},
  publisher={De Gruyter}
}

@article{bi2017uncertainty,
  title={Uncertainty quantification metrics with varying statistical information in model calibration and validation},
  author={Bi, Sifeng and Prabhu, Saurabh and Cogan, Scott and Atamturktur, Sez},
  journal={AIAA journal},
  volume={55},
  number={10},
  pages={3570--3583},
  year={2017},
  publisher={American Institute of Aeronautics and Astronautics}
}

@inproceedings{kisamori2020model,
  title={Model bridging: Connection between simulation model and neural network},
  author={Kisamori, Keiichi and Yamazaki, Keisuke and Komori, Yuto and Tokieda, Hiroshi},
  booktitle={Joint European Conference on Machine Learning and Knowledge Discovery in Databases},
  pages={389--405},
  year={2020},
  organization={Springer}
}

@article{fukumizu2013kernel,
  title={Kernel Bayes' rule: Bayesian inference with positive definite kernels},
  author={Fukumizu, Kenji and Song, Le and Gretton, Arthur},
  journal={The Journal of Machine Learning Research},
  volume={14},
  number={1},
  pages={3753--3783},
  year={2013},
  publisher={JMLR. org}
}

@article{sankararaman2013separating,
  title={Separating the contributions of variability and parameter uncertainty in probability distributions},
  author={Sankararaman, Shankar and Mahadevan, Sankaran},
  journal={Reliability Engineering \& System Safety},
  volume={112},
  pages={187--199},
  year={2013},
  publisher={Elsevier}
}

\end{document}